\documentclass[12pt]{spieman}  
\usepackage{amsmath,amsfonts,amssymb}
\usepackage{graphicx}
\usepackage{setspace}
\usepackage{tocloft}
\usepackage{textcomp}
\usepackage{multirow}
\usepackage{threeparttable}
\usepackage{booktabs}
\usepackage{caption}
\usepackage{bm}
\usepackage{algorithm,algpseudocode}
\usepackage[algo2e]{algorithm2e} 

\title{Prior-guided Diffusion Model for Cell Segmentation in Quantitative Phase Imaging}

\author[a]{Zhuchen Shao}
\author[a,b]{Mark A. Anastasio}
\author[b,c,*]{Hua Li}
\affil[a]{University of Illinois Urbana-Champaign, Department of Electrical and Computer Engineering, Urbana, Illinois, United States}
\affil[b]{University of Illinois Urbana-Champaign, Department of Bioengineering, Urbana, Illinois, United States}
\affil[c]{Washington University School of Medicine, Department of Radiation Oncology, Saint Louis, Missouri, United States}

\cftpagenumbersoff{figure}
\cftpagenumbersoff{table} 
\begin{document} 
\maketitle

    \textbf{Purpose:}  
Quantitative phase imaging (QPI) is a label-free technique that provides high-contrast images of tissues and cells without the use of chemicals or dyes. Accurate semantic segmentation of cells in QPI is essential for various biomedical applications. While DM-based segmentation has demonstrated promising results, the requirement for multiple sampling steps reduces efficiency. This study aims to enhance DM-based segmentation by introducing prior-guided content information into the starting noise, thereby minimizing inefficiencies associated with multiple sampling.

    \textbf{Approach:}  
    A prior-guided mechanism is introduced into DM-based segmentation, replacing randomly sampled starting noise with noise informed by content information. This mechanism utilizes another trained DM and DDIM inversion to incorporate content information from the to-be-segmented images into the starting noise. An evaluation method is also proposed to assess the quality of the starting noise, considering both content and distribution information.

    \textbf{Results:}  
    Extensive experiments on various QPI datasets for cell segmentation showed that the proposed method achieved superior performance in DM-based segmentation with only a single sampling. Ablation studies and visual analysis further highlighted the significance of content priors in DM-based segmentation.

    \textbf{Conclusion:}  
    The proposed method effectively leverages prior content information to improve DM-based segmentation, providing accurate results while reducing the need for multiple samplings. The findings emphasize the importance of integrating content priors into DM-based segmentation methods for optimal performance.


{\noindent \footnotesize\textbf{*}Hua Li,  \linkable{li.hua@wustl.edu} }

\begin{spacing}{2}   

\section{Introduction}
\label{sec:introduction}

Quantitative phase imaging (QPI) is a powerful, label-free approach that generates high-contrast images by measuring optical path length differences in tissue and cell samples. It is particularly effective for visualizing transparent or low-contrast biological cell structures~\cite{park2018quantitative}. 
This non-invasive approach eliminates the need for chemicals or dyes, allowing for continuous observation and monitoring of cell growth and behavior over time without interference~\cite{mir2012quantitative,vicar2019cell}. Precise cell segmentation in QPI datasets enables comprehensive examination of cell morphology and dynamics, which is crucial for a wide range of biomedical research applications~\cite{park2023artificial,jo2018quantitative,hu2022live,stringer2021cellpose}.

Traditional U-Net-based methods \cite{ronneberger2015u,yin2022u} have been widely used in cell segmentation and shown promising performance, benefiting from the ability of multi-scale feature extraction~\cite{isensee2021nnu,siddique2021u}.
In contrast, 
diffusion models (DM)~\cite{ho2020denoising, rombach2022high} generated the semantic segmentation map through a stochastic diffusion process~\cite{zbinden2023stochastic}. Given the noise randomly sampled from a Gaussian distribution, a trained DM can denoise this sampled noise to generate the segmentation mask conditioned on the input images.

As the sampling begins with randomly sampled Gaussian noise, inherent randomness may lead to performance variations in segmentation. 
Therefore, DM-based segmentation employs multiple random samplings with various starting noises to generate multiple predictions, which are then refined using ensemble learning methods like majority voting for improved accuracy~\cite{sun2023instance, chen2023berdiff, wolleb2022diffusion}.
With the help of multiple samplings and ensemble learning, DM-based segmentation has shown superior results compared to traditional segmentation methods in various medical image segmentation tasks~\cite{amit2021segdiff,sun2023instance, chen2023berdiff,wolleb2022diffusion}. 
Within the sampling process of DM, two common methods are employed: the stochastic sampling method introduced in denoising diffusion probabilistic models (DDPM) \cite{ho2020denoising}, and the deterministic sampling method introduced in denoising diffusion implicit models (DDIM) \cite{song2020denoising}. These methods are respectively referred to as ``DDPM sampling'' and ``DDIM sampling'' throughout this paper.
As the sampling process in DM is iterative and processed step-by-step, lengthy sampling times are a significant drawback of both sampling methods~\cite{song2020denoising,salimans2022progressive,zheng2023fast}. Multiple samplings further extend these sampling times, making the segmentation process more inefficient and time-consuming.

Our method proposes a novel DM-based cell segmentation approach in QPI images that reduces the need for multiple samplings and achieves better segmentation results. 
Given that QPI provides high-contrast imaging, which enhances the distinctiveness and separation of different content within the images, the proposed method integrates content information from the to-be-segmented QPI images into the starting noise used in DM sampling. In our study, this content information is referred to as the prior information of the starting noise.
In addition, an evaluation method is proposed to interpretably assess the quality of starting noise, involving both content and distribution information. Extensive experiments were conducted on various QPI datasets for cell segmentation.
Our method achieved better performance by sampling only once compared to ensemble predictions from multiple random samplings.

The remainder of the paper is organized as follows. Section II provides background information on diffusion models, DDIM sampling and its inversion process. Section III describes the proposed method. Section IV details the numerical studies, and the experimental results are presented in Section V. Finally, Section VI provides a summary and discussion of the work.

\section{Background}
\label{sec: background}
\subsection{Diffusion model}

Diffusion models (DM) are advanced generative models capable of generating high-quality images \cite{ho2020denoising,croitoru2023diffusion} and effectively learning useful semantic information \cite{shao2023semi,shao2023augdiff}. The training process of a DM involves two main steps: forward diffusion and reverse diffusion. 
In the forward diffusion process, a transition is incrementally made from the original data distribution towards a Gaussian distribution across a sequence of discrete time steps $t$, which range from $0$ to $T-1$. At each step, a predetermined amount of Gaussian noise 
$\bm{\epsilon}$ is added to the input images $\bm{x}$, progressively transforming the data as follows:
\begin{equation}
\bm{x_t}=\sqrt{\bar{\alpha}_t}\bm{x}+\sqrt{1-\bar{\alpha}_t}\bm\epsilon,
 \label{equ: diffusion_process}
\end{equation}
where $\bar{\alpha}_t$ is a hyper-parameter, $\{\bm{x_t}\}^T_{t=1}$ are noise-added images and $T$ is the maximal sampling steps. During the reverse diffusion process, the denoising U-Net denoted as $\bm{\epsilon_\theta}$ is responsible for predicting the noise in $\bm{x_t}$. The training objective is to minimize the error between the added noise $\bm{\epsilon}$ and the predicted noise $\bm{\epsilon_\theta}\left(\bm{x_t}\right)$, which can be represented as follows:
	\begin{equation}
	L_{D M}=\mathbb{E}_{\bm{x}, \bm{\epsilon} \sim \bm{\mathcal{N}}(0,1), t}\left[\left\|\bm{\epsilon}-\bm{\epsilon_\theta}\left(\bm{x_t}\right)\right\|_2^2\right].
 \label{equ: dm_loss}
	\end{equation}

Implementing a traditional DM poses challenges due to high computational costs. To address this challenge, the latent diffusion model (LDM) \cite{rombach2022high} was proposed. This method incorporates the idea of 
compression by introducing pre-trained autoencoding models to encode images into a latent space. Therefore, the training of the diffusion model in the LDM method is more efficient.
Specifically, a pre-trained autoencoding model includes an encoder $\bm{\mathcal{E}}$ and a decoder $\bm{\mathcal{D}}$. All input images $\bm{x}$ for LDM are encoded into the latent representation $\bm{z}$ by $\bm{\mathcal{E}}$. The following diffusion training process is in the latent space. The training objective can be represented as follows:
	\begin{equation}
	L_{L D M}=\mathbb{E}_{\bm{z}, \bm{\epsilon} \sim \bm{\mathcal{N}}(0,1), t}\left[\left\|\bm{\epsilon}-\bm{\epsilon_\theta}\left(\bm{z_t}\right)\right\|_2^2\right]. 
 \label{equ: ldm_loss}
	\end{equation}

\subsection{DDIM sampling and its inversion process}
\label{subsec: ddim}
 Stochastic DDPM sampling~\cite{ho2020denoising} and deterministic DDIM sampling~\cite{song2020denoising} are two commonly used sampling methods for a trained DM to generate segmentation masks. Specifically, DDPM sampling follows a Markov chain process, and the sampling process is as follows.
\begin{equation}
\bm{x_{t-1}}=\frac{1}{\sqrt{\alpha_t}}\left(\bm{x_t}-\frac{\beta_t}{\sqrt{1-\bar{\alpha}_t}} \bm{\epsilon_\theta}\left(\bm{x_t}\right)\right)+\sigma_t \bm{\epsilon},
\end{equation}
where $\bm{\epsilon_\theta}$ represents pre-trained denoising U-Net within the DM, $\bm{\epsilon}$ represents random sample Gaussian noise. The hyper-parameters $\alpha_t$, $\bar{\alpha}_t$, $\beta_t$ and $\sigma_t$ are used to determine the scales of noise. Due to the inclusion of random noise $\bm{\epsilon}$, the sampling process in DDPM is inherently stochastic.

In contrast, DDIM sampling is a deterministic process, the formulation is as follows:
\begin{equation}
\begin{aligned}
\bm{x_{t-1}}\!=\!\sqrt{\bar{\alpha}_{t-1}}\left(\frac{\bm{x_t}\!-\!\sqrt{1\!-\!\bar{\alpha}_t} \bm{\epsilon_\theta}\left(\bm{x_t}\right)}{\sqrt{\bar{\alpha}_t}}\right)\!\!+\!\!\sqrt{1\!-\!\bar{\alpha}_{t-1}} \bm{\epsilon_\theta}\left(\bm{x_t}\right)\\
=\sqrt{\frac{\bar{\alpha}_{t-1}}{\bar{\alpha}_t}} \bm{x_t}\!+\!\sqrt{\bar{\alpha}_{t-1}}\left(\sqrt{\frac{1}{\bar{\alpha}_{t-1}}\!-\!1}\!-\!\sqrt{\frac{1}{\bar{\alpha}_t}\!-\!1}\right) \bm{\epsilon_\theta}\left(\bm{x_t}\right),
\end{aligned}
\label{equ: ddim}
\end{equation}
where $\bm{\epsilon_\theta}$ represents pre-trained denoising U-Net within the DM, while the hyper-parameter $\bar{\alpha}_t$ determines the scales of the noise. The DDIM process does not involve random noise \( \bm{\epsilon} \). This process is defined as a deterministic sampling method and can be reformulated to derive its corresponding ordinary differential equation (ODE)~\cite{song2020denoising}. Consequently, numerical approximation methods, including the Euler method, are capable of approximating the inversion process of DDIM sampling.
By using the Euler method with a sufficient number of discrete time steps, the DDIM inversion is effectively deterministic.

The deterministic characteristic of DDIM sampling and DDIM inverse sampling ensures consistent transformation between input images and their corresponding noise. To facilitate the recovery of the original image from noise, the DDIM inverse sampling method preserves important content information during its noise addition process. Inspired by the ability to retain content information, DDIM inversion is employed in areas such as image translation~\cite{su2022dual,tumanyan2023plug} and image editing~\cite{miyake2023negative} for extracting content information of input images.

\begin{figure*}[h]
    \centerline{\includegraphics[width=\linewidth]{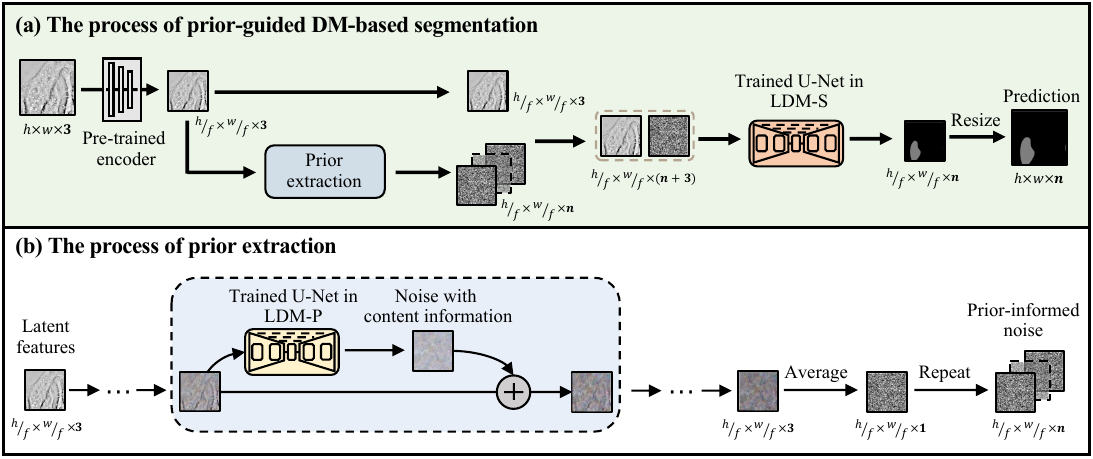}}
\caption{\textbf{Proposed prior-guided DM-based segmentation (PG-DiffSeg) method.} (a) The process of prior-guided DM-based segmentation, 
(b) The process of prior information extraction. Pre-trained LDM-P was employed to generate content-informed starting noise directly from a given image through DDIM inversion process. 
This obtained noise, informed by the image's content and following a Gaussian distribution, will be input to the LDM-S for segmentation. 
    }
    \label{figure_overview}
\end{figure*}

\section{Proposed framework}
\subsection{Overview of the proposed method}
As shown in Fig.~\ref{figure_overview}, the proposed method is named the prior-guided DM-based segmentation (PG-DiffSeg). 
Leveraging the high-contrast content information in QPI images, this method incorporates content information into the starting noise of the sampling process, enabling DM-based segmentation to achieve precise results without multiple samplings and ensemble learning.
Specifically, this method utilizes two latent diffusion models (LDMs)~\cite{rombach2022high}: one LDM for prior extraction (LDM-P) and another for segmentation (LDM-S). 
By exploiting the deterministic DDIM inversion process, LDM-P transforms to-be-segmented images into corresponding noise, which conforms to a Gaussian distribution while retaining content information. Instead of relying on randomly sampled noise, LDM-S utilizes the prior-informed noise extracted by LDM-P as the starting noise to generate the segmentation mask.
Starting from the prior-informed noise better supports LDM-S in stably generating a precise segmentation mask.

\subsection{Latent diffusion model for prior extraction (LDM-P)} 
\label{diffusion_prior}
To generate Gaussian noise while retaining content information in images, a latent diffusion model (LDM) with the DDIM inverse sampling process is employed to transform input images into their corresponding noises.
The LDM used in this process is named LDM-P. 
Specifically, the training of LDM-P involves both the forward and reverse diffusion processes in an unsupervised manner. For efficient training, the input images are transformed into latent representations $\bm{z}$ using a pre-trained encoder $\bm{\mathcal{E}}$~\cite{rombach2022high}. During the forward diffusion process, random noise is introduced to $\bm{z}$, as illustrated in Eq.\ref{equ: diffusion_process}. Subsequently, during the reverse diffusion process, LDM-P predicts the added noise by minimizing the training loss defined in Eq.~\ref{equ: ldm_loss}. Given to-be-segmented images, the trained LDM-P model converts the input latent representation into prior-informed noise for DM-based segmentation described below.

\subsection{Latent diffusion model for segmentation (LDM-S)}
 
 

Given the prior-informed noise generated by LDM-P, another LDM is employed to predict the segmentation mask of the to-be-segmented image.
In our study, this LDM for segmentation tasks is referred to as LDM-S.
Training LDM-S involves both images and segmentation masks and follows the supervised learning process. The paired inputs are first pre-processed.
The to-be-segmented images $\bm{x}$ are transferred to latent representation $\bm{z}$ using the pre-trained encoder $\bm{\mathcal{E}}$~\cite{rombach2022high}. 
The segmentation mask $\bm{y}$ is initially converted into multiple binary segmentation masks through one-hot encoding and then resized to match the dimensions of the latent representation $\bm{z}$.
The training of LDM-S involves both forward and reverse diffusion processes. In the forward diffusion, Gaussian noise is added to the pre-processed segmentation mask \( \bm{y} \), as illustrated in Eq.~\ref{equ: diffusion_process}. 
During the reverse diffusion, with the condition of the to-be-segmented image, a denoising U-Net \( \bm{\epsilon_\theta} \) is utilized to predict the added noise on the segmentation masks \( \bm{y_t} \). 
The training objective aims to minimize the loss function, defined as the squared error between the added noise \( \bm\epsilon \) and the noise predicted by the model, \( \bm{\epsilon_\theta}(\bm{y_t}, \bm{z}) \):
\begin{equation}
L=\mathbb{E}_{\bm{y}, \bm{z}, \bm{\epsilon} \sim \mathcal{N}(0,1), t}\left[\left\|\bm{\epsilon}-\bm{\epsilon_\theta}\left(\bm{y_t}, \bm{z}\right)\right\|_2^2\right].
\end{equation}

The trained LDM-S model predicts the segmentation mask using the prior-informed noise $\bm{\epsilon_p}$ as the starting noise, which is generated by the trained LDM-P on the given to-be-segmented image.
The segmentation in LDM-S requires the dimensions of the starting noise to align with those of the segmentation mask. The prior-informed noise $\bm{\epsilon_p}$ obtained from LDM-P has a dimension of $\bm{\epsilon_p}\in\mathbb{R}^{h/f \times w/f \times 3}$, where $h \times w$ corresponds to the original image size, and $f$ signifies the down-sampling factor. Averaging and replication operations are performed on $\bm{\epsilon_p}$ to preprocess it into $\bm{\epsilon_p}\in\mathbb{R}^{h/f \times w/f \times n}$, where $n$ represents the number of classes in the segmentation mask.

Using this pre-processed starting noise, the trained LDM-S employs the DDIM sampling process (Eq.~\ref{equ: ddim}) to generate segmentation mask predictions. Subsequently, a resizing operation is applied to convert the prediction from the latent space size to the image space.  
The testing process of LDM-S is presented in Algorithm~\ref{alg:collaboration}.

\begin{algorithm}[h]
    \small
    \caption{Testing of the LDM-S}
    \label{alg:collaboration}
    \KwIn{
        Input image $\bm{x}$, informed noise $\bm{\epsilon_p}$ from Section~\ref{diffusion_prior}. Pre-trained encoder $\bm{\mathcal{E}}$ from \cite{rombach2022high}. LDM-S is referred to as $\bm{\mathcal{D_S}}$. $h \times w$ indicates the original image dimensions, while $h_f \times w_f$ represents the reduced image dimensions in the latent space.
    }
    \KwOut{Predicted segmentation result $\bm{\hat{y}}$.}
        
        {{\textbf{1)}} Pre-processing the prior-informed noise and input image.}\\
        $\bm{\epsilon_p^a} = \operatorname{Average}(\bm{\epsilon_p})$, $\bm{\epsilon_p^a}\in\mathbb{R}^{h_f\times w_f\times 1}$ \\
        $\bm{\epsilon_p^r} = \operatorname{Repeat}(\bm{\epsilon_p^a})$, $\bm{\epsilon_p^r}\in\mathbb{R}^{h_f\times w_f\times n}$ {//} $n$ is the number of class\\
        $\bm{z} = \bm{\mathcal{E}}\left(\bm{x}\right)$, $\bm{z}\in\mathbb{R}^{h_f\times w_f\times 3}$\\
        {{\textbf{2)}} Using pre-trained LDM-S to predict segmentation mask.}\\
        $\bm{\hat{y}} = \operatorname{Resize}\left(\bm{\mathcal{D_S}}\left(\bm{\epsilon_p^r},\bm{z} \right)\right)$, $\bm{\hat{y}}\in\mathbb{R}^{h\times w\times n}$ 
\end{algorithm}

\section{Numerical Studies}
\subsection{Datasets}
\label{subsec: datasets}
Two QPI datasets for cell viability assessment study \cite{hu2022live}, HeLa and CHO-2, were utilized to evaluate model performance on semantic segmentation, aiming to distinguish between live and dead cells. For simplicity, CHO-2 was referred to as CHO in this study.
The image size is 832$\times$832 pixels, each accompanied by a corresponding segmentation mask.
The HeLa dataset contains 1,199 images, while the CHO dataset comprises 2,051 images. The original images were subdivided into smaller sizes of 256$\times$256 and split into training, validation and testing datasets. Details about the data can be found in Table~\ref{tab: dataset}.

\begin{table}[]
\centering
    \caption{ 
Statistical overview of datasets after pre-processing
    }
    \label{tab: dataset}
\begin{threeparttable}
\setlength{\tabcolsep}{6mm}
    \begin{tabular}{c|cc}
        \toprule
        
                   & \multicolumn{1}{c|}{HeLa}    & {CHO}                \\ \midrule
\#image            & \multicolumn{1}{c|}{19,184}   & 22,816                 \\
\#train   & \multicolumn{1}{c|}{14,400}    & 16,416               \\
\#validation & \multicolumn{1}{c|}{3,200}      & 3,200                \\
\#test       & \multicolumn{1}{c|}{1,584}      & 3,200                \\
image size         & \multicolumn{1}{c|}{256$\times$256} & 256$\times$256              \\ 
\#classes         & \multicolumn{1}{c|}{3}  & 3              \\ 
\bottomrule
\end{tabular}
  \end{threeparttable}
\end{table}

\subsection{Model architecture}
The LDM-P and LDM-S were used for prior extraction and segmentation, respectively. The denoising U-Net within both LDMs follows the architecture designed in LDM~\cite{rombach2022high}, with the same number of U-Net layers, ResNet-structured blocks, and self-attention blocks. The model architecture for comparative methods is described in Sec.~\ref{subsec: comparative methods}.
The LDM-P prioritizes learning overall content information with a larger receptive field. Accordingly, it features larger hidden layers with dimensions of 224, 448, 672, and 896, along with higher down-sampling factors where the spatial resolution of input features to attention modules is set to 8, 16, and 32.
Conversely, the LDM-S emphasizes learning structure and local information, necessitating larger spatial resolution in self-attention blocks. Accordingly, the spatial resolution of input features to attention modules is set to 16, 32, and 64. To reduce the computational burden, it incorporates a smaller hidden layer with dimensions of 128, 256, 384, and 512. The parameters for LDM-P and LDM-S are 274M and 84.6M, respectively. 

\subsection{Implementation details}

In the experiments, the parameters predefined by LDM~\cite{rombach2022high} were utilized. The configuration involved a maximum diffusion step of 1000, and the pre-trained weight named ``vq-f4" was selected for the encoder $\bm{\mathcal{E}}$ to transform the size of input images from 256 to 64.
In the training process, the LDM-P used a batch size of 20 and a learning rate of 2e-06, while the LDM-S utilized a batch size of 10 and a learning rate of 2e-06.
The training for both LDMs involved standard data augmentation techniques, including rotation, horizontal and vertical flipping, as well as random adjustments to brightness and contrast. During testing, the LDM-P utilized a 100-step DDIM inversion to transform input images into prior-informed noise. The LDM-S employed a 200-step DDIM sampling (Eq.~\ref{equ: ddim}) in generating the segmentation mask. 

\subsection{Comparison methods}
\label{subsec: comparative methods}
Five state-of-the-art segmentation methods were implemented for comparison. Three of them are deterministic segmentation methods, while the other two are diffusion model-based segmentation methods using multiple sampling and ensemble learning.

\subsubsection{Deterministic segmentation methods} 
The proposed method was compared to three deterministic segmentation methods, including \textbf{1)}~ResNet-U-Net~\cite{he2016deep,ronneberger2015u} and \textbf{2)}~Efficient-U-Net~\cite{tan2019efficientnet,ronneberger2015u}, as well as \textbf{3)}~DeepLabv3~\cite{chen2017rethinking}. Both ResNet-U-Net and DeepLabv3 used resnet101~\cite{he2016deep} as encoder, and Efficient-U-Net used efficientnetb4~\cite{tan2019efficientnet} as encoder. For the decoder part, both ResNet-U-Net and Efficient-U-Net used U-Net decoder. Different from U-Net decoder, the decoder in DeepLabv3 used atrous spatial pyramid pooling (ASPP) module for capturing multi-scale information. The model parameters for ResNet-U-Net, Efficient-U-Net, and DeepLabv3 were 51.5M, 20.2M, and 58.6M, respectively. To ensure a fair comparison, all the comparative methods were trained from scratch.

For the training of these three methods, a batch size of 50 and a learning rate of 1e-04 with the Lookahead+Radam optimizer \cite{zhang2019lookahead} were employed. The training process also employed the same data augmentation as our method. Dice loss~\cite{sudre2017generalised} was employed for model training.

\subsubsection{DM-based segmentation methods}
The proposed method was also compared with two approaches: \textbf{1)} DM-based segmentation based on multiple samplings~\cite{amit2021segdiff} and \textbf{2)} DM-based segmentation guided with a pre-trained DINO~\cite{caron2021emerging} named CCDM method~\cite{zbinden2023stochastic}. Randomly sampled starting noise is used in both methods. Additionally, CCDM integrated the DINO to extract features from the input images and then combined them with the features extracted in the denoising U-Net, thereby incorporating image content information into the segmentation process.

The training of traditional diffusion model-based segmentation was similar to the proposed PG-DiffSeg. For training CCDM, a batch size of 8 and a learning rate of 2e-06 were used. The pretrained weight for DINO~\cite{caron2021emerging} was applied with the ``dino\_vits8" setting, and the dimension of the extracted features was 384. These features were concatenated with the input features of the fourth U-Net block in LDM. The model parameters for the LDM-implemented CCDM were 108M.

\subsection{Evaluation metrics}
\label{subsec: evaluation metrics}
Mean Intersection over Union (mIoU) and F1 score (F1) were used to evaluate the segmentation performance. Higher values for these metrics indicate better performance.
The Structural Similarity Index (SSIM) was used to assess the content information retained in the starting noise for sampling.
SSIM takes into account structure, texture, and luminance information, ranging from -1 to 1, with 1 indicating perfect similarity.
Additionally, the Probability Density Function (PDF) of the starting noise is estimated using the Gaussian Kernel Density Estimation technique.
Kullback-Leibler Divergence (KLD) is applied to determine the difference between the estimated PDF of the starting noise and that of the standard Gaussian distribution. The value of KLD ranges from 0 to positive infinity, with 0 indicating a standard Gaussian distribution.
Both content information assessment and distribution assessment aim to analyze the superiority of the prior-informed noise in retaining content image information while still conforming to a Gaussian distribution.
With three-fold cross-validation, all evaluation metrics included the mean and standard deviation (SD).
Independent samples t-tests were conducted to identify significant differences in the results and were reported with \textit{p}-values.

\section{Results}
\subsection{Cell segmentation results}
Experiments were conducted on two label-free imaging datasets to assess the effectiveness of the method in stratifying and contouring live and dead cells. Section~\ref{subsec: datasets} provided the dataset details, while Sections~\ref{subsec: comparative methods} and~\ref{subsec: evaluation metrics} discussed the comparative methods and evaluation metrics, respectively. The results are summarized in Table \ref{tab: single}. It is worth noting that ``Random$\times1$" represents DM-based segmentation with only one sampling, while ``Random$\times$3/$\times$5" and ``CCDM$\times$3/$\times$5" represent 3 or 5 times of multiple sampling. Majority voting is employed to obtain ensemble predictions from these multiple predictions. Our proposed PG-DiffSeg is denoted as ``Prior". 

\textbf{1)} Compared to two U-Net-based segmentation methods, the DM-based method with single sampling (``Random$\times$1") can achieve comparable performance. By using multiple samplings in the DM-based method, the segmentation performance increased compared to ``Random$\times$1" and also achieved better results than all the deterministic segmentation methods. This observation confirms that multiple sampling plays an important role in improving the performance of DM-based segmentation methods.

\textbf{2)} Nonetheless, increasing the number of multiple sampling does not always guarantee performance improvement. For example, from ``Random$\times$1'' to ``Random$\times$3'', the mIoU and F1 increased 3.5\% and 3.1\% in HeLa dataset, respectively. However, from ``Random$\times$3'' to ``Random$\times$5'', the mIoU and F1 remained the same performance. Same trend was also observed on CCDM.  
The possible reason is after a certain point, additional samplings may not provide significant new information or may even introduce noise into the segmentation process, resulting in a negative effect in performance.
Compared to ``Random", the ``CCDM" method showed higher performance on the HeLa dataset but not the CHO dataset. The distinction between "Random" and "CCDM" lies in the use of a pre-trained DINO model with a large amount of natural images to guide the diffusion training process in CCDM. The content information extracted by DINO had a positive impact on the HeLa dataset, whereas it had a negative effect on the CHO dataset. This difference may be attributed to the more complex structure of cell images within the CHO dataset, which significantly differs from the characteristics of the natural images used for pre-training DINO, leading to the ineffectiveness of the CCDM method.

\textbf{3)} Our proposed prior-guided method for DM-based segmentation significantly enhanced performance in terms of mIoU and F1 over both the HeLa and CHO datasets. Furthermore, by incorporating content information specific to the to-be-segmented images into the starting noise, only a single sampling was required to achieve optimal performance.

\begin{table}[h]
\centering
    \caption{
        Results for cell segmentation on the QPI datasets
    }
    \label{tab: single}
\begin{threeparttable}
    \begin{tabular}{c|cc|cc}
        \toprule
\multirow{2}{*}{}            & \multicolumn{2}{c|}{\underline{HeLa}}              & \multicolumn{2}{c}{\underline{CHO}}       \\
                             & mIoU       & F1                    & mIoU       & F1             \\ \midrule
ResNet-U-Net                         & 0.589 (0.026)          & 0.671 (0.019)         & 0.584 (0.047)         & 0.652 (0.046)         \\
Efficient-U-Net               & 0.592 (0.006)         & 0.675 (0.011)         & 0.570 (0.045)         & 0.645 (0.041)         \\
DeepLabv3               & {0.618} (0.047)          & {0.696} (0.031)                 & 0.627 (0.064)          & 0.692 (0.060)          \\\midrule
Random$\times$1 & 0.582 (0.045)          & 0.657 (0.030)                   & 0.584 (0.094)         & 0.646 (0.090)         \\
Random$\times$3 & 0.617 (0.067)          & 0.688 (0.051)                    & 0.640 (0.086)          & 0.698 (0.082)          \\
Random$\times$5 & {0.618} (0.050)          & 0.688 (0.034)                    & \underline{0.656} (0.081)          & \underline{0.712} (0.078)          \\ \midrule
CCDM$\times$1 &  0.602 (0.058)          & 0.669 (0.045)                   & 0.519 (0.049)         & 0.581 (0.047)         \\
CCDM$\times$3 & \underline{0.637} (0.096)          & \underline{0.698} (0.083)                    & 0.601 (0.079)          & 0.661 (0.077)          \\
CCDM$\times$5 & {0.630} (0.078)          & 0.690 (0.066)                    & 0.619 (0.080)          & 0.678 (0.078)          \\ \midrule
Prior (Ours)     & \textbf{0.653} (0.061) & \textbf{0.725} (0.045)  & \textbf{0.661} (0.070) & \textbf{0.715} (0.070) \\ 
\bottomrule
\end{tabular}
          \begin{tablenotes}
\item[1] The results were reported as ``average performance (standard deviation)" due to the cross validation. The highest values were presented in bold, and the second-highest values were underlined.
    \end{tablenotes}
  \end{threeparttable}
\end{table}

\subsection{Ablation study}

\subsubsection{{Effects of DM-based prior extraction network}} 
The LDM-P serves as the prior extraction network in the proposed method. To evaluate the effectiveness of this trainable module (``LDM-P'' in Table~\ref{tab: ablation_pretrained}), two different starting noises were compared in DM-based segmentation, involving ``Random'' and ``Forward Diff``. 
In the ``Random'' approach, the starting noise is randomly sampled from a Gaussian distribution. The starting noise in the ``Forward Diff.'' approach is generated following the forward diffusion process (Eq.~\ref{equ: diffusion_process}) to the input images directly. Three starting noises were generated by using 0, 300, and 600 forward diffusion steps, respectively. Increasing the forward diffusion steps introduces more Gaussian noise, consequently preserving less content information.
Additionally, in the ``LDM-P'' method, three different starting noises were generated using different DDIM inversion steps of 50, 100, and 200, respectively. A larger DDIM inversion step indicates a smaller stride in DDIM sampling and a more precise sampling process. 

Table~\ref{tab: ablation_pretrained} illustrates the results for different approaches on two datasets of HeLa and CHO. According to the metrics of mIoU and F1, it is evident that on both datasets, incorporating content prior information into the starting noise improved segmentation accuracy compared to the random sampling setting, regardless of whether it was generated by LDM-P or the ``Forward Diff.'' method. In addition, utilizing DDIM inversion yielded better average performance compared to the non-trainable forward diffusion method, with the average mIoU improving by at least 0.9\% in the HeLa dataset and by 6.3\% in the CHO dataset.

To explain how the starting noise affects the segmentation performance, SSIM and KLD metrics were used to evaluate the quality of the starting noise. The results shown in Table~\ref{tab: ablation_pretrained} validated that our proposed prior extraction method struck a better balance by preserving more content information in the starting noise and closely aligning with the Gaussian distribution. In the ``Random'' setting, the starting noise consisted of pure Gaussian noise without any content information, as indicated by both SSIM and KLD of 0. In ``Forward Diff. 0'', starting noise is set as the image itself, meaning it retained the most content information but also deviated the most from the standard Gaussian distribution. Gaussian noise was introduced to the to-be-segmented image through 300 and 600 forward diffusion steps, respectively in ``Forward Diff. 300'' and ``Forward Diff. 600''. Due to the minimal noise addition, the KLD metric in ``Forward Diff. 300'' was found to be 10 times higher compared to other methods, which negatively impacted its performance. Additionally, ``Forward Diff. 300'' had lower KLD by applying more forward diffusion steps, but it also lost much content information in this process, leading to suboptimal segmentation results. In addition to ``Forward Diff. 0'' and ``Forward Diff. 300'', both non-trainable and trainable methods that incorporated content prior information exhibited significantly larger SSIM metrics than random sampling (\textit{p}-value\textless 0.05) and showed small KLD metrics, indicating a closer alignment with the standard Gaussian distribution. Interestingly, when applying different DDIM inversion steps in our LDM-P, there was not a significant difference observed, indicating the robustness in selecting this hyper-parameter.

\begin{table*}[h]
\centering
    \caption{
        Ablation study for DM-based prior extraction network.
    }
\label{tab: ablation_pretrained}
\begin{threeparttable}
\begin{tabular}{c|cccc}
\toprule
\multirow{2}{*}{} & \multicolumn{4}{c}{\underline{HeLa}}                              \\
                  & mIoU$\uparrow$  & F1$\uparrow$                & SSIM$\uparrow$  & KLD$\downarrow$  \\ \midrule
Random            & \multicolumn{1}{c}{0.582(0.045)} & 0.657(0.030) & 0 & 0                   \\ \midrule
Forward Diff. 0     & \multicolumn{1}{c}{0.636(0.089)} & 0.704(0.072) & 1 & inf  \\
Forward Diff. 300     & \multicolumn{1}{c}{0.650(0.058)} & 0.721(0.044) & \textbf{0.655}(0.169) & 0.019(0.001)  \\
Forward Diff. 600     & \multicolumn{1}{c}{0.644(0.064)} & 0.715(0.052) & 0.215(0.063) & \textbf{0.001}(0.000) \\ \midrule
LDM-P 50      & \multicolumn{1}{c}{\textbf{0.656}(0.062)} & \textbf{0.728}(0.046) & 0.229(0.040) & \textbf{0.001}(0.000) \\
LDM-P 100     & \multicolumn{1}{c}{\underline{0.653}(0.061)} & \underline{0.725}(0.045) & \underline{0.230}(0.043) & \textbf{0.001}(0.000) \\
LDM-P 200     & \multicolumn{1}{c}{\underline{0.653}(0.061)} & \underline{0.725}(0.045) & \underline{0.230}(0.044) & \textbf{0.001}(0.000)  \\ \bottomrule

\toprule
\multirow{2}{*}{}                       & \multicolumn{4}{c}{\underline{CHO}}                          \\      
                  & mIoU$\uparrow$  & F1$\uparrow$                & SSIM$\uparrow$  & KLD$\downarrow$        \\ \midrule
Random              & \multicolumn{1}{c}{0.584(0.093)} & 0.646(0.090) & 0 & 0           \\ \midrule
Forward Diff. 0     & \multicolumn{1}{c}{0.620(0.016)} & {0.677(0.014)} & 1 & inf \\
Forward Diff. 300     & \multicolumn{1}{c}{0.639(0.070)} & 0.698(0.068) & \textbf{0.609}(0.001) & 0.009(0.000) \\
Forward Diff. 600     & \multicolumn{1}{c}{0.597(0.102)} & 0.657(0.099) & 0.164(0.000) & 0.002(0.000) \\ \midrule
LDM-P 50       & \multicolumn{1}{c}{0.660(0.071)} & 0.715(0.069) & 0.192(0.016) & \textbf{0.001}(0.000) \\
LDM-P 100     & \multicolumn{1}{c}{\textbf{0.661}(0.070)} & \textbf{0.716}(0.069) & {0.194}(0.017) & \textbf{0.001}(0.000) \\
LDM-P 200     & \multicolumn{1}{c}{\textbf{0.661}(0.070)} & \textbf{0.716}(0.069) & \underline{0.197}(0.019) & \textbf{0.001}(0.000) \\ \bottomrule
\end{tabular}
          \begin{tablenotes}
\item[1] The results were reported as ``average performance (standard deviation)" due to the cross validation. The highest values were presented in bold, and the second-highest values were underlined.
    \end{tablenotes}
  \end{threeparttable}
\end{table*}

\subsubsection{{Effects of DDIM inversion method}} 
The trained LDM-P employed DDIM inversion to extract content information from to-be-segmented images, integrating this with starting noise that conforms to a Gaussian distribution. This study explores the impact of DDIM inversion on content information extraction compared to two alternative methods. In the first method, features extracted from the self-attention module of the LDM-P contain segmentation-relevant content information~\cite{tumanyan2023plug}. However, the starting noise generated during this feature extraction may not conform to standard Gaussian noise, potentially reducing the efficiency of DM-based segmentation.
In the second method, content information is extracted from input images using a U-Net block~\cite{shao2023semi} and is then utilized as the starting noise for the subsequent segmentation process. Both methods rely on features extracted from the fifth block of the denoising LDM-P, with 400 noise addition steps during the forward diffusion process.
Additionally, the "Random" method, which indicates DM-based segmentation without DDIM inversion, was also compared.
mIoU, F1, SSIM, and KLD were used as the evaluation metrics.
The results are summarized in Table~\ref{tab: ablation_pretrained2}.

Based on the mIoU and F1 results, it was evident that the features extracted from the self-attention modules performed better than those from the U-Net block. This finding indicates that the long-distance information extraction capability of the self-attention module can produce better starting noise than the U-Net block. However, neither feature extraction method achieved satisfactory results in the KLD metric, showing a significant deviation from the standard Gaussian distribution. Although these methods yielded higher SSIM metrics in the CHO dataset, the poor performance in distribution information negatively impacted segmentation results.  In contrast, DDIM inversion proved to be the optimal approach for using trained LDM-P to obtain starting noise for following DM-based segmentation tasks.

\begin{table*}[h]
\centering
    \caption{
        Ablation study for DDIM inversion method.
    }
\label{tab: ablation_pretrained2}
\begin{threeparttable}
\begin{tabular}{c|cccc}
\toprule
\multirow{2}{*}{} & \multicolumn{4}{c}{\underline{HeLa}}                        \\
                  & mIoU$\uparrow$  & F1$\uparrow$    
                  & SSIM$\uparrow$ & KLD$\downarrow$         \\ \midrule
Random / w/o DDIM inversion            & \multicolumn{1}{c}{0.582(0.045)} & 0.657(0.030)  & 0 & 0                  \\ \midrule
Feature Extraction (U-Net block)     & \multicolumn{1}{c}{0.514(0.171)} & 0.569(0.172)  & \underline{0.104}(0.075) & \underline{0.106}(0.025)  \\
Feature Extraction (Self-attention)     & \multicolumn{1}{c}{\underline{0.619}(0.079)} & \underline{0.689}(0.058)  & 0.087(0.052) & 0.127(0.027)  \\
DDIM inversion     & \multicolumn{1}{c}{\textbf{0.653}(0.061)} & \textbf{0.725}(0.045) & \textbf{0.230}(0.043) & \textbf{0.001}(0.000)  \\ \bottomrule

\toprule
\multirow{2}{*}{}                        & \multicolumn{4}{c}{\underline{CHO}}                          \\
                  & mIoU$\uparrow$  & F1$\uparrow$  & SSIM$\uparrow$                & KLD$\downarrow$        \\ \midrule
Random / w/o DDIM inversion               & \multicolumn{1}{c}{0.584(0.093)} & 0.646(0.090)  & 0  & 0           \\ \midrule
Feature Extraction (U-Net block)     & \multicolumn{1}{c}{0.531(0.062)} & {0.578(0.063)} & \underline{0.274}(0.010) & \underline{0.050}(0.004) \\
Feature Extraction (Self-attention)     & \multicolumn{1}{c}{\underline{0.649}(0.055)} & \underline{0.705}(0.055) & \textbf{0.383}(0.037) & 0.078(0.014) \\
DDIM inversion     & \multicolumn{1}{c}{\textbf{0.661}(0.070)} & \textbf{0.716}(0.069) & 0.194(0.017) & \textbf{0.001}(0.000) \\ \bottomrule
\end{tabular}
          \begin{tablenotes}
\item[1] The results were reported as ``average performance (standard deviation)" due to the cross validation. The highest values were presented in bold, and the second-highest values were underlined.
    \end{tablenotes}
  \end{threeparttable}
\end{table*}

\subsection{Visualization}
\begin{figure*}[!t]
    \centerline{\includegraphics[width=\linewidth]{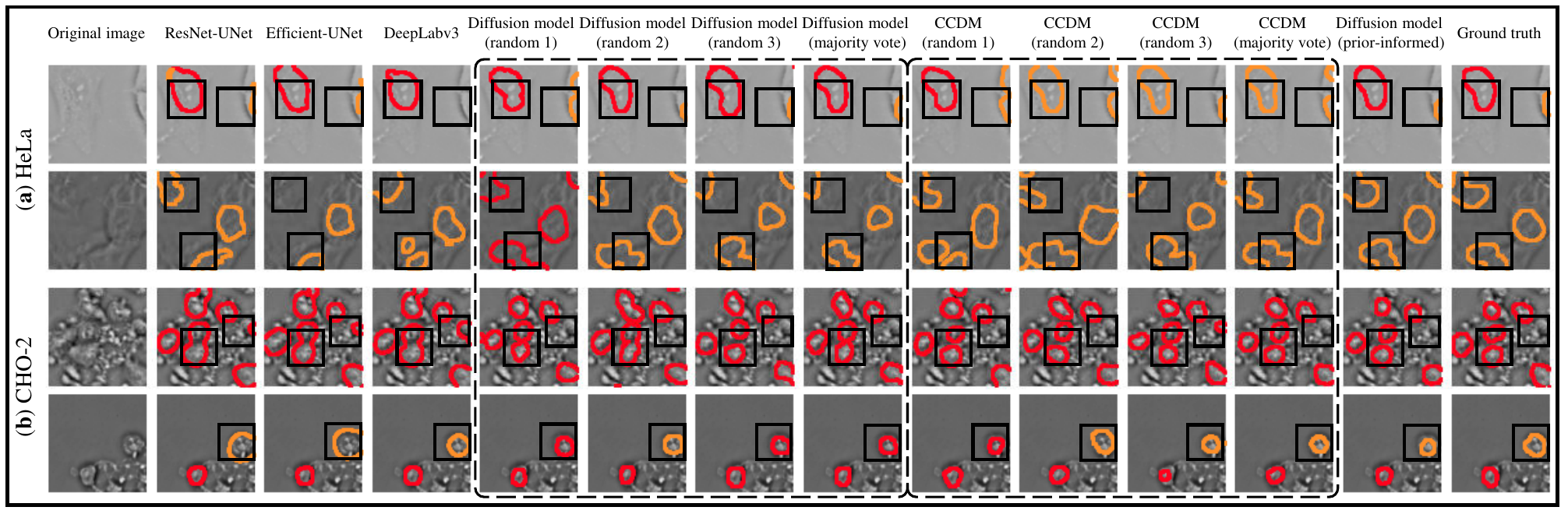}}
\caption{\textbf{Visualization analysis.} Visualization results are presented for all methods across two single annotation tasks. The dashed block highlights that these results are specific to the random sampling-based method. Live cells are represented in red, while dead cells are shown in orange. The black square highlights areas where the proposed method performs better than other methods.
    }
    \label{figure_visualization}
\end{figure*}

\begin{figure}[!t]
    \centerline{\includegraphics[width=\linewidth]{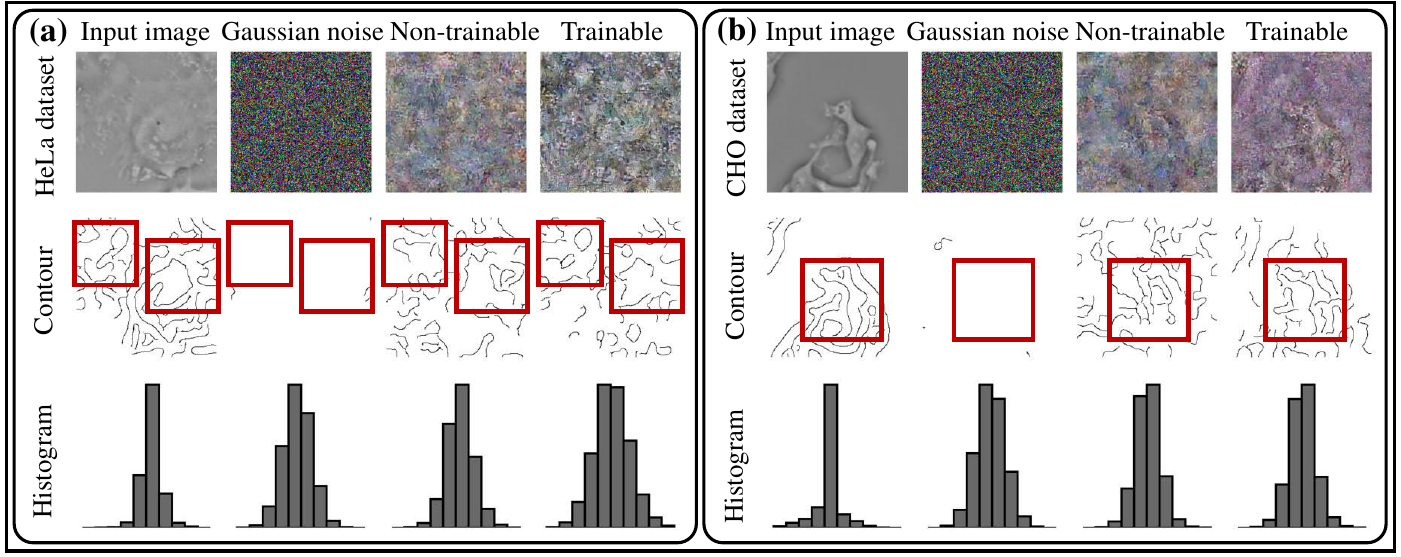}}
\caption{\textbf{Visualization analysis for the content prior.} The canny edge detector is utilized to obtain contour information, while histograms are employed to analyze distribution information. The red box highlights the specified area.} 
    \label{figure_visualization_interpretability}
\end{figure}

Visualization was used to demonstrate the effectiveness of the proposed method. The results are shown in Fig.~\ref{figure_visualization}. Additionally, the visualization of content prior is examined in Fig.~\ref{figure_visualization_interpretability}.

In Fig.~\ref{figure_visualization}, the HeLa and CHO datasets were obtained using label-free imaging, focusing on distinguishing live cells, dead cells, and the background through semantic segmentation. It is clear from the dashed box that adding random noise to DM-based segmentation generates multiple different predictions. Using ensemble learning, like the majority vote, can improve overall prediction performance. However, these multiple predictions faced the problem of unpredictable variance, potentially misleading ensemble learning. Our method accurately predicted results using prior information with just one sample. Compared to deterministic methods such as U-Net, our approach also achieved better segmentation accuracy.

In Fig.~\ref{figure_visualization_interpretability}, a comparison among random sampling (Gaussian distribution), the non-trainable method (forward diffusion), and the trainable method (DDIM inversion) is presented. The canny edge detector~\cite{ding2001canny} is employed to analyze content information, while contrast enhancement followed by Gaussian blur is used to preprocess the starting noise. Both the HeLa and CHO datasets are utilized for content information and distribution analysis.
From the histogram analyses, it can be observed that both the starting noise obtained by non-trainable and trainable methods follow a Gaussian distribution similar to Gaussian noise. Additionally, the contour information is retained in the starting noise for these two methods, while Gaussian noise does not retain any content information due to the randomness. Moreover, the red box indicates that the trainable method can retain more contour information than the non-trainable method, which is significantly important for segmentation. 

\section{Discussion and conclusion}
Utilizing the high-contrast content information in QPI images, this study introduces a novel DM-based segmentation framework called PG-DiffSeg, which integrates a prior extraction network to enhance DM-based segmentation in both segmentation accuracy and speed. By utilizing DDIM inversion to extract content prior information from the to-be-segmented images, this method addresses the issue of lacking content information in randomly sampled starting noise. Furthermore, this paper proposes a unique approach involving SSIM and KLD metrics to assess the effectiveness of starting noise, emphasizing the importance of content prior and data distribution information. Extensive experiments were conducted on two QPI datasets for cell segmentation. Our method achieved satisfactory results with just one sampling. The ablation studies confirm the effectiveness of the proposed approaches and the robustness of the hyper-parameter selection.

Future research on the prior-guided DM-based segmentation framework could expand in multiple directions. One potential direction is to improve the prior extraction network by integrating classification-related prior information into the starting noise. This adjustment may potentially enhance the performance of semantic segmentation, particularly when dealing with a larger number of classes. Additionally, in the LDM-based segmentation method, replacing Gaussian noise with Bernoulli noise could be advantageous, given that the generation target is a one-hot encoding segmentation mask. Furthermore, conducting additional experiments on a diverse range of datasets would facilitate comprehensive testing of the model's generalization capabilities.

\subsection*{Disclosures}
The authors declare that they have no known competing financial interests or personal relationships that could have appeared to influence the work reported in this paper.

\subsection* {Code, Data, and Materials Availability} 
Code and data will be made publicly available upon acceptance of the paper.



\bibliography{report}   
\bibliographystyle{spiejour}   





\end{spacing}
\end{document}